\documentclass[10pt,conference]{IEEEtran}

\usepackage{amsmath,amssymb,amsfonts}
\usepackage{graphicx}
\usepackage{tikz}
\usepackage{cite}
\usepackage{hyperref}
\usepackage{algorithm,algpseudocode}
\usepackage[font=footnotesize]{caption}
\usepackage{subcaption}
\title{Parameters Fixing Strategy for Quantum Approximate Optimization Algorithm}

\author{%
	\IEEEauthorblockN{Xinwei Lee}
	\IEEEauthorblockA{\textit{Graduate School of Systems and Information Engineering} \\
		\textit{University of Tsukuba}\\
		Tsukuba, Ibaraki, JAPAN \\
		xwlee@cavelab.cs.tsukuba.ac.jp}
	\and
	\IEEEauthorblockN{Yoshiyuki Saito}
	\IEEEauthorblockA{\textit{Graduate School of Computer Science and Engineering} \\
		\textit{University of Aizu} \\
		Aizu-Wakamatsu, Fukushima, JAPAN \\
		m5241141@u-aizu.ac.jp}
	\and
	\IEEEauthorblockN{Dongsheng Cai}
	\IEEEauthorblockA{\textit{Faculty of Engineering, Information and Systems} \\
		\textit{University of Tsukuba}\\
		Tsukuba, Ibaraki, JAPAN \\
		cai@cs.tsukuba.ac.jp}
	\and
	\IEEEauthorblockN{Nobuyoshi Asai}
	\IEEEauthorblockA{\textit{School of Computer Science and Engineering} \\
		\textit{University of Aizu} \\
		Aizu-Wakamatsu, Fukushima, JAPAN \\
		nasai@u-aizu.ac.jp}
}

\IEEEoverridecommandlockouts
\IEEEpubid{\begin{minipage}{\textwidth}\ \\[1cm] \copyright2021~IEEE. Personal use of this material is permitted. Permission from IEEE must be obtained for all other uses, in any current or future media, including %
		reprinting/republishing this material for advertising or promotional purposes, creating new collective works, for resale or redistribution to servers or lists, or reuse of any copyrighted component of %
		this work in other works.\end{minipage}}

\newcommand{\ER}{Erd\"{o}s-R\'{e}nyi }
\newcommand{\ket}[1]{| #1 \rangle}
\newcommand{\bra}[1]{\langle #1 |}
\newcommand{\vgamma}{\vec{\gamma}}
\newcommand{\vbeta}{\vec{\beta}}
\newcommand{\Optvec}{(\vgamma^*,\vbeta^*)}

\newcommand{\Gbvec}{(\vgamma,\vbeta)}
\newcommand{\Gb}[1]{(\gamma_{#1},\beta_{#1})}
\newcommand{\sidecaption}[1]
{\raisebox{\abovecaptionskip}{\begin{subfigure}[t]{1.6em}
			\caption[singlelinecheck=off]{}
			\label{#1}
	\end{subfigure}}\ignorespaces}

\renewcommand{\figurename}{Fig.}

\begin{document}
\maketitle

\begin{abstract}
	The quantum approximate optimization algorithm (QAOA) has numerous promising applications in solving the combinatorial optimization problems on near-term Noisy Intermediate Scalable Quantum (NISQ) devices.
	QAOA has a quantum-classical hybrid structure. Its quantum part consists of a parameterized alternating operator ansatz, and its classical part comprises an optimization algorithm, which optimizes the
	parameters to maximize the expectation value of the problem Hamiltonian. This expectation value depends highly on the parameters, this implies that a set of good parameters leads to an accurate solution.
	However, at large circuit depth of QAOA, it is difficult to achieve global optimization due to the multiple occurrences of local minima or maxima. In this paper, we propose a parameters fixing strategy which
	gives high approximation ratio on average, even at large circuit depths, by initializing QAOA with the optimal parameters obtained from the previous depths. We test our strategy on the Max-cut problem of
	certain classes of graphs such as the 3-regular	graphs and the \ER graphs.
\end{abstract}

\section{Introduction}
Since the introduction of QAOA by Farhi et al.~\cite{farhi2014quantum}, it is widely known for its efficiency and universality in solving combinatorial optimization problems on quantum
computers~\cite{lloyd2018quantum,farhi2019quantum}. Among the problems, the Max-cut
problem is heavily studied for its simple formulation and its deep relation to the Ising model~\cite{AuriacOptimCuts}.
There exist some classes of graphs which the Max-cut problem can be solved analytically, such as the bipartite graphs, the 2-regular (ring) graphs and the fully-connected graphs. However, for the other classes
of graphs there is no analytical solution, and they are NP-hard~\cite{Kar72}.

Although QAOA does not give the exact solution to the Max-cut problem, the algorithm provides a heuristic approach for the problem.
For classical heuristics, the Goemans-Williamson algorithm (widely known as GW algorithm) is able to achieve the approximation ratio of at least 0.878 on the Max-cut problem through semidefinite
programming~\cite{GW1995}. Farhi et al. stated in~\cite{farhi2014quantum} that QAOA is able to achieve the approximation ratio of at least 0.6924 on 3-regular graphs for $p=1$, where $p$ is the depth of the
quantum circuit for QAOA. Recently, Wurtz et al.~\cite{Wurtz_2021} extends the result to $p>1$, with the lower bound of the approximation ratio as 0.7559 for $p=2$ and 0.7924 for $p=3$, for 3-regular graphs.
It is obvious that the lower bound for the approximation ratio increases as $p$ increase, and will surpass the lower bound for the GW algorithm for large enough $p$. This agrees with~\cite{farhi2014quantum} that
as $p\rightarrow\infty$, the approximation ratio obtainable from QAOA approaches 1, in which we obtain the true solution to the problem.

However, several hurdles continue to exist in the practical application of QAOA. One of them is obtaining a ``good'' solution at large circuit depths.
Solving large problems (graphs with large number of nodes) often requires deep circuits. However, due to the existence of local maxima on the hypersurface of the
expectation function, local optimizers tend to be trapped inside one of the local maximum, whilst the desired solution are the global maxima. Many researches, for example~\cite{Guerreschi_2019,niu2019,Willsch_2020,Moussa_2020,	leo2020,Cook2020TheQA,Daniel_bangbang,sack2021quantum,multistart},
have stated this problem as the limitation of QAOA at large circuit depths. Some researches~\cite{brandao2018fixed,leo2020,Cook2020TheQA,Daniel_bangbang,sack2021quantum,multistart,Alam2020ML}
discussed the strategies to improve the performance of QAOA, which will be discussed later. It is now clear that the performance of QAOA heavily depends on its initial parameters,
whether it converges to a local maximum or a global maximum, and hence its performance. This motivates our work to introduce a strategy that aims to choose the QAOA parameters such that they
yield better performance overall. \IEEEpubidadjcol

In this paper, we discuss a straightforward, yet practically effective, parameters fixing strategy to improve the average performance of QAOA at large circuit depths. First, using simulation results, we show that
with random initial angles, QAOA will have a poor performance on average at large circuit depths and large problems. With our strategy applied, the average performances on the same problems are improved.
We choose our problems from the widely studied 3-regular graphs and the \ER random graphs to verify the capability of our strategy in solving those problem instances.

\section{QAOA for Max-cut}
\begin{figure}[t]
	\centering
	\includegraphics{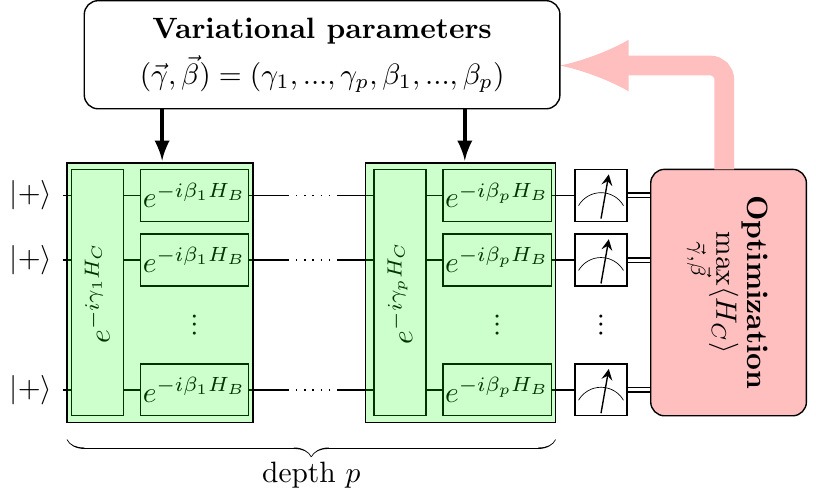}
	\caption{The routine of the Quantum Approximate Optimization Algorithm (QAOA).}
	\label{fig:qaoa_circ}
\end{figure}

QAOA originates from the Quantum Adiabatic Algorithm (QAA)~\cite{farhi:qaa}, which focuses on evolving the initial Hamiltonian $H_B$ to the problem Hamiltonian $H_C$, satisfying
\begin{equation}
	\tilde{H}(s) = (1-s)H_B + sH_C,
	\label{eqn:qaa}
\end{equation}
where $s(t)\rightarrow 1$ as $t\rightarrow\infty$. The evolution in Eq.~(\ref{eqn:qaa}) is then discretized, which results in QAOA. In QAOA, the alternating unitary operators involving $H_B$ and $H_C$
are applied to the initial state to simulate the evolution of the system in Eq.~(\ref{eqn:qaa}):
\begin{equation}
	|\psi_p(\vgamma,\vbeta)\rangle = e^{-i\beta_p H_B}e^{-i\gamma_p H_C}\ldots e^{-i\beta_1 H_B}e^{-i\gamma_1 H_C}|+\rangle^{\bigotimes n},
	\label{eqn:ansatz}
\end{equation}
where $\vgamma = (\gamma_1, \gamma_2, \ldots, \gamma_p)$ and $\vbeta = (\beta_1, \beta_2, \ldots, \beta_p)$ are the $2p$ variational parameters, with $\gamma_i\in[0,2\pi)$ and $\beta_i\in[0,\pi)$.
$\ket{+}^{\bigotimes n}$ corresponds to $n$ qubits in the ground state of the Pauli-$X$ basis.
For the Max-cut problem of a graph $G = (V, E)$, $H_C$ and $H_B$ are given as
\begin{align}
	H_C & = \frac{1}{2}\sum_{(j, k)\in E}(I - Z_j Z_k), \label{eqn:hc}\\
	H_B & = \sum_{j\in V}X_j \label{eqn:hb}.
\end{align}
$X_j$ and $Z_j$ are the Pauli operators acting on the $j$-th qubit. After applying the operators as in Eq.~(\ref{eqn:ansatz}), we calculate the expectation of the operator $H_C$ with respect to the ansatz state
$|\psi_p(\vgamma,\vbeta)\rangle$:
\begin{equation}
	F_p(\vgamma,\vbeta) = \langle\psi_p(\vgamma,\vbeta)|H_C|\psi_p(\vgamma,\vbeta)\rangle.
	\label{eqn:fp}
\end{equation}
Since Eq.~(\ref{eqn:fp}) is parameterized by $\vgamma$ and $\vbeta$, we can use a classical optimization algorithm to search for the angles which maximize $F_p$:
\begin{equation}
	(\vgamma^*,\vbeta^*) = \arg\max_{\vgamma,\vbeta} F_p(\vgamma,\vbeta).
\end{equation}
Fig.~\ref{fig:qaoa_circ} shows the schematic diagram of such a routine for QAOA. First, some initial parameters $(\vgamma, \vbeta)$ is passed into
the quantum circuit of QAOA, which is characterized by Eq.~(\ref{eqn:ansatz}), (\ref{eqn:hc}) and (\ref{eqn:hb}). The results of the measurement are then used to calculate
$F_p$ from (\ref{eqn:fp}), and the value of $F_p$ is optimized using a classical optimizer to obtain a better $(\vgamma, \vbeta)$ to be passed into the quantum circuit again.
This continues until the optimizer converges and the optimal parameters $(\vgamma^*, \vbeta^*)$ is obtained.
The \emph{approximation ratio} $\alpha$ is defined as
\begin{equation}
	\alpha = \frac{F_p(\vgamma^*,\vbeta^*)}{C_{\text{max}}},
	\label{eqn:alpha}
\end{equation}
where $C_{\text{max}}$ is the maximum cut value for the graph. The approximation ratio indicates how near the solution given by QAOA is to the true solution, ranging from 0 to 1, with
0 the furthest and 1 the nearest to the true solution. It will be the performance measure throughout this paper as we aim to approximate the solution of the Max-cut problem as accurate as possible.

As the circuit depth $p$ increases, $\max F_p(\vgamma,\vbeta)$ also increases, and it will approach $C_{\max}$ as $p\rightarrow\infty$, thus the approximation ratio $\alpha$ will approach 1~\cite{farhi2014quantum}.
However, according to our numerical experiments, the optimizers does not necessarily succeed in finding the maximum value of $F_p$ as $p$ gets larger. Apparently, which parameters are chosen
as the initial points to the optimizer plays an important role in whether the optimizer converges to a global maximum. Hence, we would prefer to have ``good'' initial points in order for the optimizer to
converge to the global maximum.

\section{Related Work}
While many strategies of choosing good parameters for QAOA are introduced and discussed, there is no strong evidence that which
strategy is more effective than the others. Some~\cite{leo2020,Cook2020TheQA,sack2021quantum,Crooks2018PerformanceOT} found out that there is a linear relationship between the parameters and the circuit depths, which
resembles the linear annealing scheme. As shown in their work, the optimal parameters $\gamma_i$ increases linearly and $\beta_i$ decreases linearly, where $i$ is the index of the parameters in $\Gbvec$.
However, besides the linearly increasing and decreasing parameters, the landscape of the expectation function also has multiple maximum points, which arguments are some sets of parameters
that do not follow the linear pattern. In fact, in most cases of our simulations, the optimal parameters we obtain do not follow the linear pattern.

Brandao et al.~\cite{brandao2018fixed} found out that for the Max-cut problem, the parameters for the typical instances of 3-regular graphs concentrate. This results in similar expectation function
for different 3-regular graphs regardless the number of nodes, as long as the structure of the graph is the same. They leverage this property to reuse the optimal parameters obtained from optimizing
smaller graphs in the optimization of larger graphs, which they call it the leapfrogging strategy. They state that this allows them to find good solutions using QAOA with fewer calls to the quantum computer.
Their strategy also applies to \ER graphs with edge probability $3/(n-1)$, as they have the expected graph degree of 3.

Zhou et al.~\cite{leo2020} introduces two strategies for optimizing the variational parameters. First, they discover the linear relationship between the parameters and the circuit depth $p$
in solving the Max-cut problem on 16-vertex unweighted 3-regular graphs using QAOA. They use the linear interpolation method to extend the optimal parameters at depth $p$ to make a guess about
the parameters at $p+1$. They call this the INTERP heuristic strategy. The second strategy is the FOURIER heuristic strategy, which uses the Discrete Sine/Cosine Transform to determine the parameters at $p+1$
from the Fourier amplitudes at $p$. They test the two strategies against the random initialization of the parameters on 40 instances of 16-vertex unweighted and weighted 3-regular graphs,
and their strategies outperform the random initialization QAOA after $p\geq 5$.

Shaydulin et al.~\cite{multistart} uses a multistart approach to improve the performance of QAOA on graph clustering problems. They combine two optimization algorithms, namely APOSMM~\cite{aposmm1,aposmm2} and
BOBYQA~\cite{bobyqa}.
APOSMM is an optimization algorithm which coordinates multiple local optimization runs to identify better local optima. They embed the classical local optimizer BOBYQA in APOSMM, which they call the APOSMM+BOBYQA,
to improve the QAOA performance. They apply this strategy to solve the modularity maximization community detection problem with graphs between 10 and 12 vertices. They have shown, using simulations,
that the APOSMM+BOBYQA method performs better than typical optimizers such as COBYLA~\cite{cobyla}, NELDER-MEAD~\cite{neldermead}, and also the BOBYQA alone.
They also apply the aforementioned leapfrogging strategy in~\cite{brandao2018fixed} to reuse optimal parameters as initial points for different graphs.

A more recent work by Sack et al.~\cite{sack2021quantum} introduce the initialization based on the Trotterized quantum annealing (TQA) protocol to prevent the convergence to false local minima. This parameter
initialization method is inspired by the connection between simulated quantum annealing and QAOA. They suggest the initialization of the $i$-th parameter in depth $p$ as
$\gamma_i = (i/p)\Delta t$ and $\beta_i = (1-i/p)\Delta t$, which is derived from the discretization of quantum annealing protocol using Suzuki-Trotter decomposition. They also found out that there exist
optimal time steps $\Delta t$ which helps in the convergence to a minima that is very close to the global minima. They apply their strategy to solve the Max-cut problem for 12-vertices over 50 random graphs.
Others like~\cite{Alam2020ML} use machine learning techniques to predict the initial parameters for higher depth QAOA. After the models are trained, they demonstrate an average improvement
of 44.9\% in run-time across various local optimization algorithms such as L-BFGS-B, NELDER-MEAD, SLSQP and COBYLA.

\section{Numerical Experiment}
In this paper, the performances of two initialization methods of QAOA are compared: the random initialization and the parameters fixing strategy. We use QAOA simulation to solve the Max-cut problem
on the 3-regular graphs and the \ER random graphs with edge probability of 0.5 with different problem sizes (number of nodes).
The simulation of the quantum circuits in QAOA is done by using the Qiskit Aer simulator and the problem graphs are generated using the NetworkX package. We solve the Max-cut problem for the 3-regular graphs
with the number of nodes $n = 6, 8, 10, 12, 16$. We also solve for the \ER graphs with $n = 6, 7, 8, 9, 10$.
The graph instances are chosen randomly so that they do not have biased structures.

With the parameters initialization methods aside, our common simulation procedure is as follows: first, we pass a set of initial parameters $(\vgamma_{\text{init}}, \vbeta_{\text{init}})$ into the
quantum circuit shown in Fig.~\ref{fig:qaoa_circ}. We then measure the quantum circuit to obtain the probability distribution of the ansatz state. We calculate the expectation value as
\begin{align}
	F(\vgamma, \vbeta) & = \bra{\psi(\vgamma,\vbeta)} H_C \ket{\psi(\vgamma,\vbeta)} \\
	& = \sum_i \lambda_i p_i,
\end{align}
where $\lambda_i$ is the eigenvalue (cut-value) of the problem Hamiltonian $H_C$, and $p_i$ is the probability of measuring the eigenstate correspond to $\lambda_i$. The expectation value is then passed
into the optimizer for a better set of parameters. This cycle continues until the optimizer terminates and we have the maximized expectation $F(\vgamma^*, \vbeta^*)$ with its optimal parameters
$(\vgamma^*, \vbeta^*)$. Hence, we can calculate the approximation ratio $\alpha$ for this set of parameters from Eq.~(\ref{eqn:alpha}). In all of our simulations, we use the NELDER-MEAD~\cite{neldermead} optimizer
provided by the SciPy package, with the settings of 1000 maximum allowed function evaluations and the absolute error tolerance of $10^{-4}$. As for the metric to evaluate the performances of the
initialization methods, we use the \emph{mean approximation ratio} obtained from the QAOA simulations initialized by different initial parameters, for a fixed QAOA circuit depth $p$. Additionally,
we investigate how the distribution of the approximation ratios changes with $p$.

\begin{figure*}[!ht]
	\centering
	\begin{subfigure}[t]{0.45\textwidth}
		\centering
		\sidecaption{subfig:a}
		\raisebox{-\height}{\includegraphics[width=\textwidth]{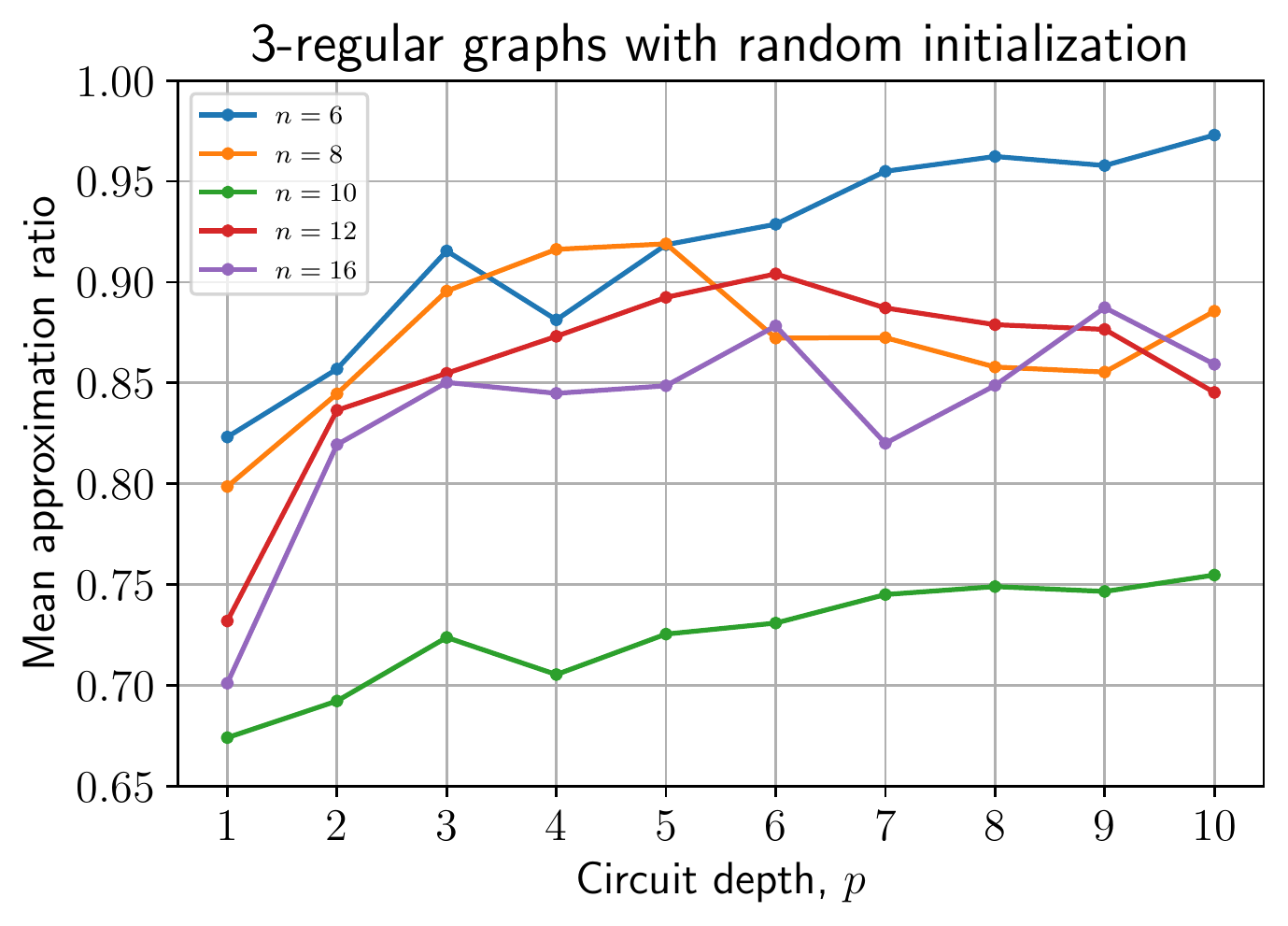}}
	\end{subfigure}
	\hspace{.7cm}
	\begin{subfigure}[t]{0.45\textwidth}
		\centering
		\sidecaption{subfig:b}
		\raisebox{-\height}{\includegraphics[width=\textwidth]{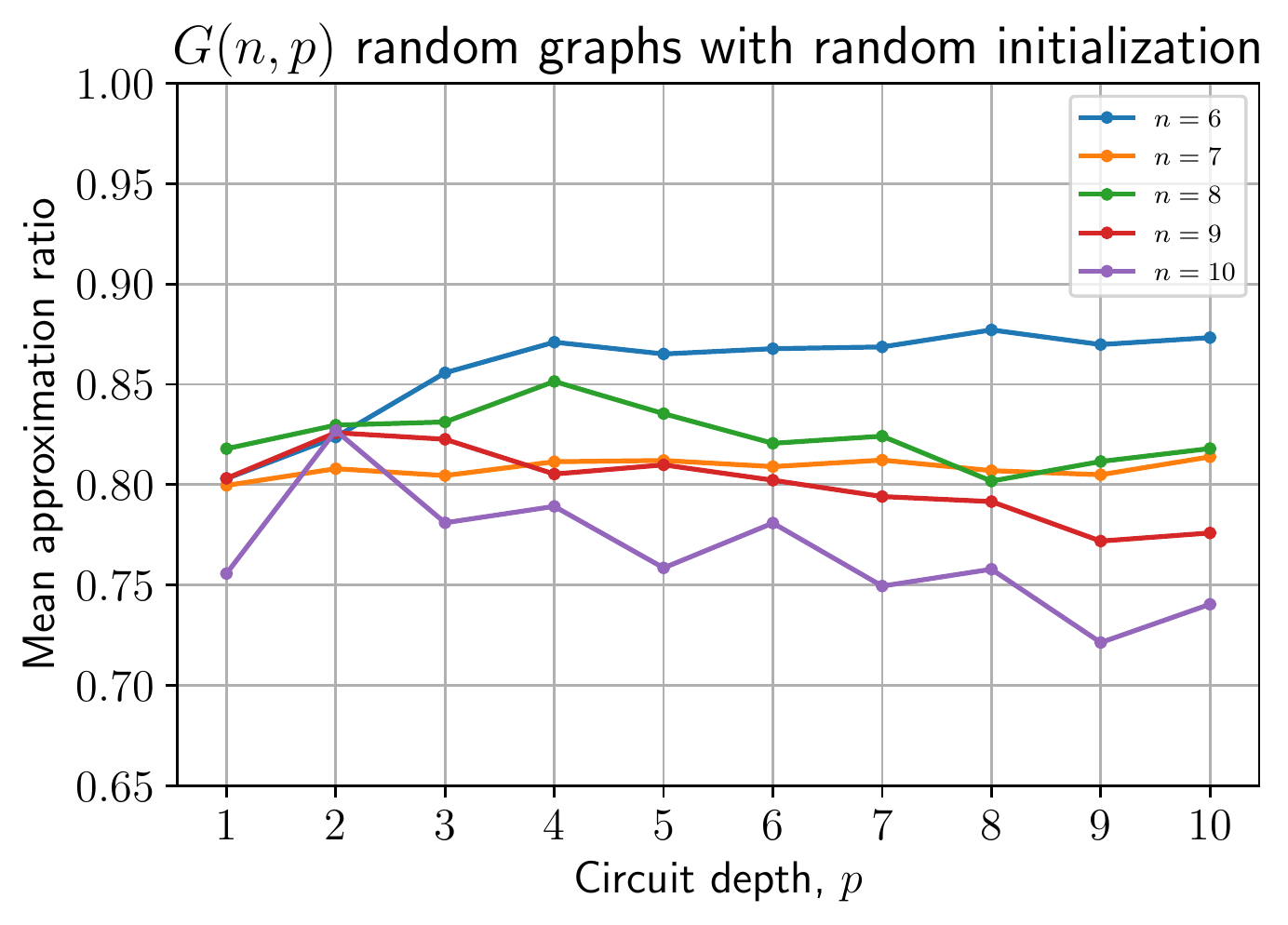}}
	\end{subfigure}
	\begin{subfigure}[t]{0.45\textwidth}
		\centering
		\sidecaption{subfig:c}
		\raisebox{-\height}{\includegraphics[width=\textwidth]{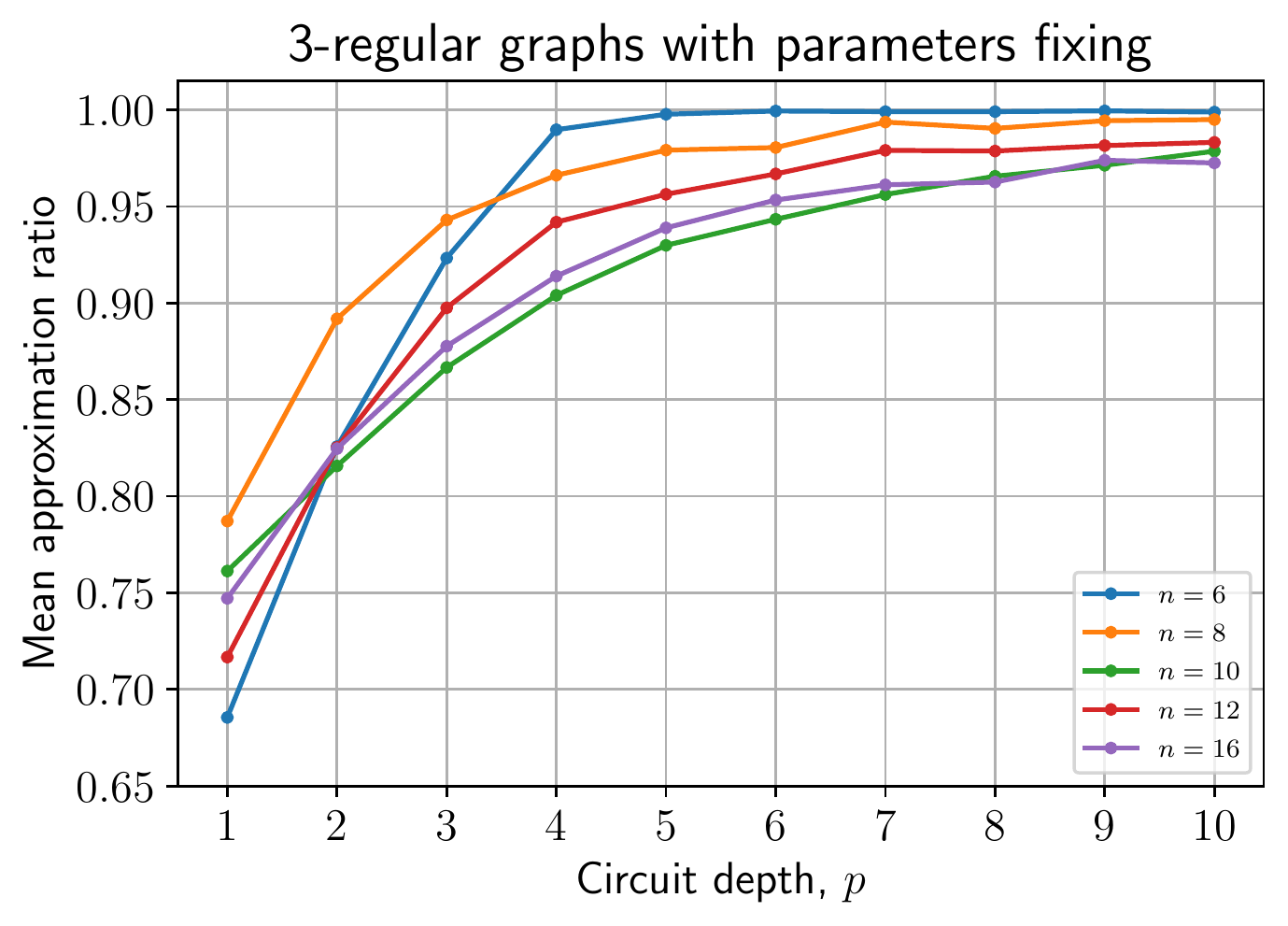}}
	\end{subfigure}
	\hspace{.7cm}
	\begin{subfigure}[t]{0.45\textwidth}
		\centering
		\sidecaption{subfig:d}
		\raisebox{-\height}{\includegraphics[width=\textwidth]{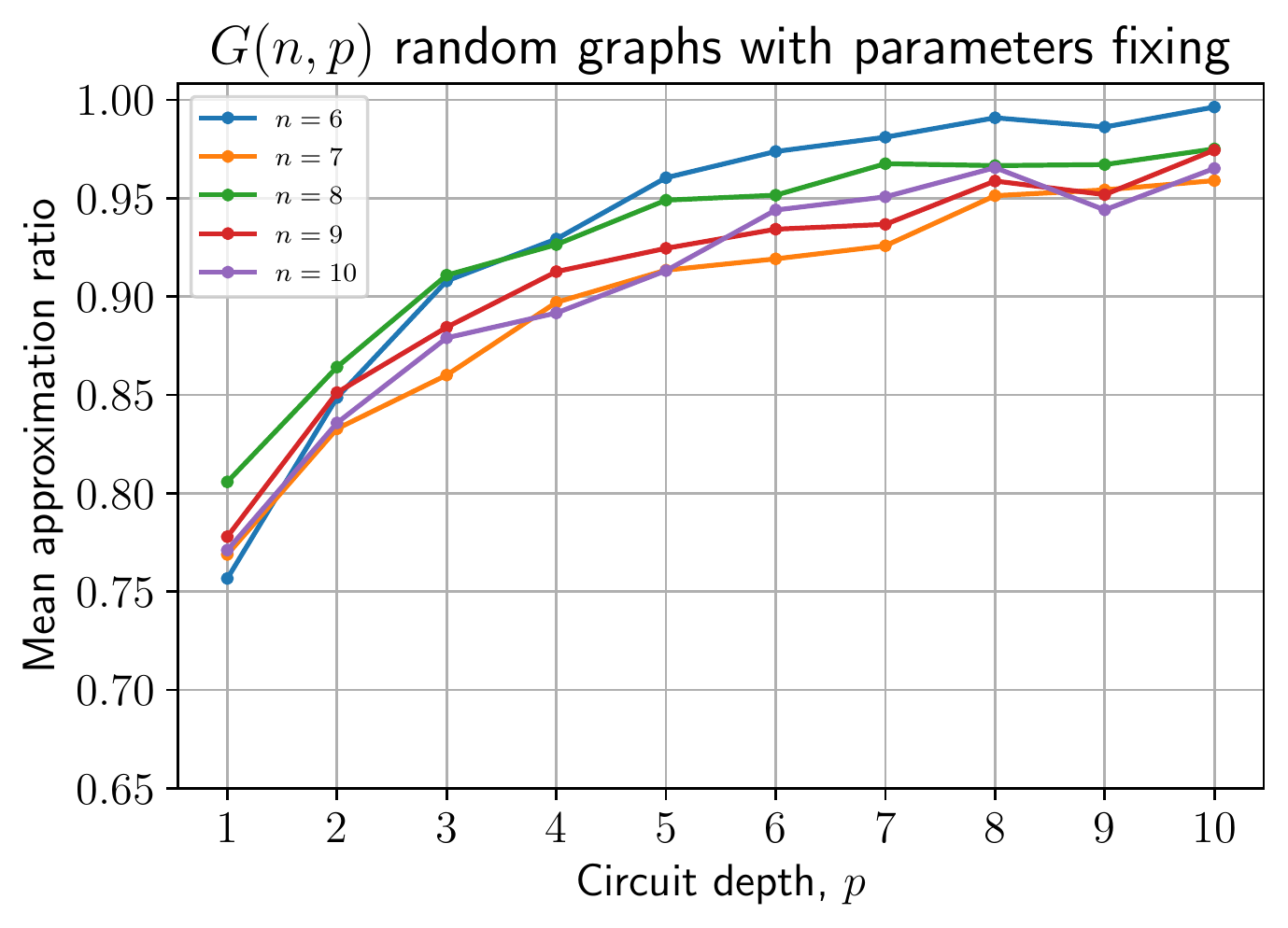}}
	\end{subfigure}
	\caption{The comparison of the results produced by applying different parameters initialization methods. The Max-cut instances solved are the 3-regular graphs ($n=6,8,10,12,16$) and %
	the \ER (or sometimes $G(n,p)$) graphs ($n=6,7,8,9,10$) with edge probability of 0.5. The variation of the mean approximation ratio against the circuit depth is plotted. (a) 3-regular graphs with %
	random initialization, (b) \ER graphs with random initialization, (c) 3-regular graphs with parameters fixing strategy, (d) \ER graphs with parameters fixing strategy.}
	\label{fig:qaoa-results}
\end{figure*}

\subsection{Random Initialization Method}
For the random initialization method, 20 sets of random parameters $(\vgamma, \vbeta)$ are passed into QAOA for each $p$ and we take the average of the 20 approximation ratios output by QAOA.
Then, we increase $p$ and repeat. The same procedure is done to different graph instances. This is the most naive method used when solving problems using QAOA. The output data will be the reference
to which after our strategy is applied. Fig.~\ref{fig:qaoa-results}(a) and (b) show the results of the QAOA simulations using random initialization method. We can observe that for 3-regular graphs,
the mean approximation ratio given
by 20 random initial parameters does not increase with the circuit depth $p$ overall. For small number of nodes like $n=6$, the mean approximation ratio increases with $p$ and is able to achieve mean $\alpha>0.95$
for $p\geq 7$. However, as the number of nodes increases ($n\geq 8$), the mean $\alpha$ only increase for small depths. As the depth gets larger ($p\geq 5$), there is a decrease in the mean $\alpha$ and most
of them do not achieve $\alpha>0.95$ like the 6-node graphs did, even the depth is increased until $p=10$. For some $n$, the standard deviations of $\alpha$ increase as $p$ increases. This
implies that at larger $p$, there exist some initial parameters which converged to our desired real maxima, giving high approximation ratio. On the other hand, there are also some initial parameters which converged
to some false local maxima, giving low approximation ratio which dragged down the mean $\alpha$. For the \ER graphs, similar trend is observed for $n\geq 7$.

\subsection{Parameters Fixing Strategy}
At large circuit depth of QAOA, the optimizer does not necessarily find the desired global maxima. Inspired by the linear interpolation strategies in~\cite{leo2020,Cook2020TheQA},
we attempt to make better choices of parameters by extending them from depth $p$ to $p+1$. Starting from $p=1$, we solve the problem with 20 sets of random parameters. As the optimizer behaves well in
shallow depth of QAOA, we entrust it to find our desired solution at $p=1$. We then select the parameters with the highest approximation ratio and treat it as the optimal parameters for $p=1$:
$(\gamma^*_1, \beta^*_1)$. We fix this set of parameters and insert another pair of random $\Gb{2}$ and pass $(\gamma^*_1, \gamma_2, \beta^*_1, \beta_2)$ as the initial parameters into solving $p=2$.
Again, to make sure we find the true global maxima, instead of using one random pair of $\Gb{i}$, we use 20 random pairs of $\Gb{i}$ and select the one which gives us the highest approximation ratio. Hence,
we obtain the optimal parameters at $p=2$ as $(\gamma^*_1, \gamma^*_2, \beta^*_1, \beta^*_2)$. This procedure is repeated until $p=10$. This is summarized in Algorithm~\ref{alg:pf}.
We call this the \emph{parameters fixing} strategy.


\begin{algorithm}[t]
	\caption{Parameters Fixing}\label{alg:pf}
	\begin{algorithmic}[1]
		\For{$q=1...p$}\Comment{Circuit depth $p$}
			\For{$k=1...n$}\Comment{No. of trials $n$}
				\State $\gamma_q\leftarrow\text{rand}(0,2\pi)$
				\State $\beta_q\leftarrow\text{rand}(0,\pi)$
				\If{$q=1$}
					\State $\Gbvec\leftarrow\Gb{1}$
					\Else
					\State $\Gbvec\leftarrow(\gamma^*_1,...,\gamma^*_{q-1},\gamma_q,\beta^*_1,...,\beta^*_{q-1},\beta_q)$
					\Statex \Comment{New parameters $(\gamma_q,\beta_q)$ are inserted.}
				\EndIf
				\State Initialize QAOA with $\Gbvec$ and optimize to find $F_q\Optvec_k$.
			\EndFor
			\State $F_q\Optvec\leftarrow\max_k F_q\Optvec_k$
			\State $\Optvec\leftarrow\arg\max_k F_q\Optvec_k$
			\Statex \Comment{$\Optvec = (\gamma^*_1,...,\gamma^*_q,\beta^*_1,...,\beta^*_q)$}
		\EndFor
		\State \textbf{Output:} $\Optvec$ and $F_p\Optvec$.
	\end{algorithmic}
\end{algorithm}

\begin{figure}[t]
	\centering
	\includegraphics[width=0.45\textwidth]{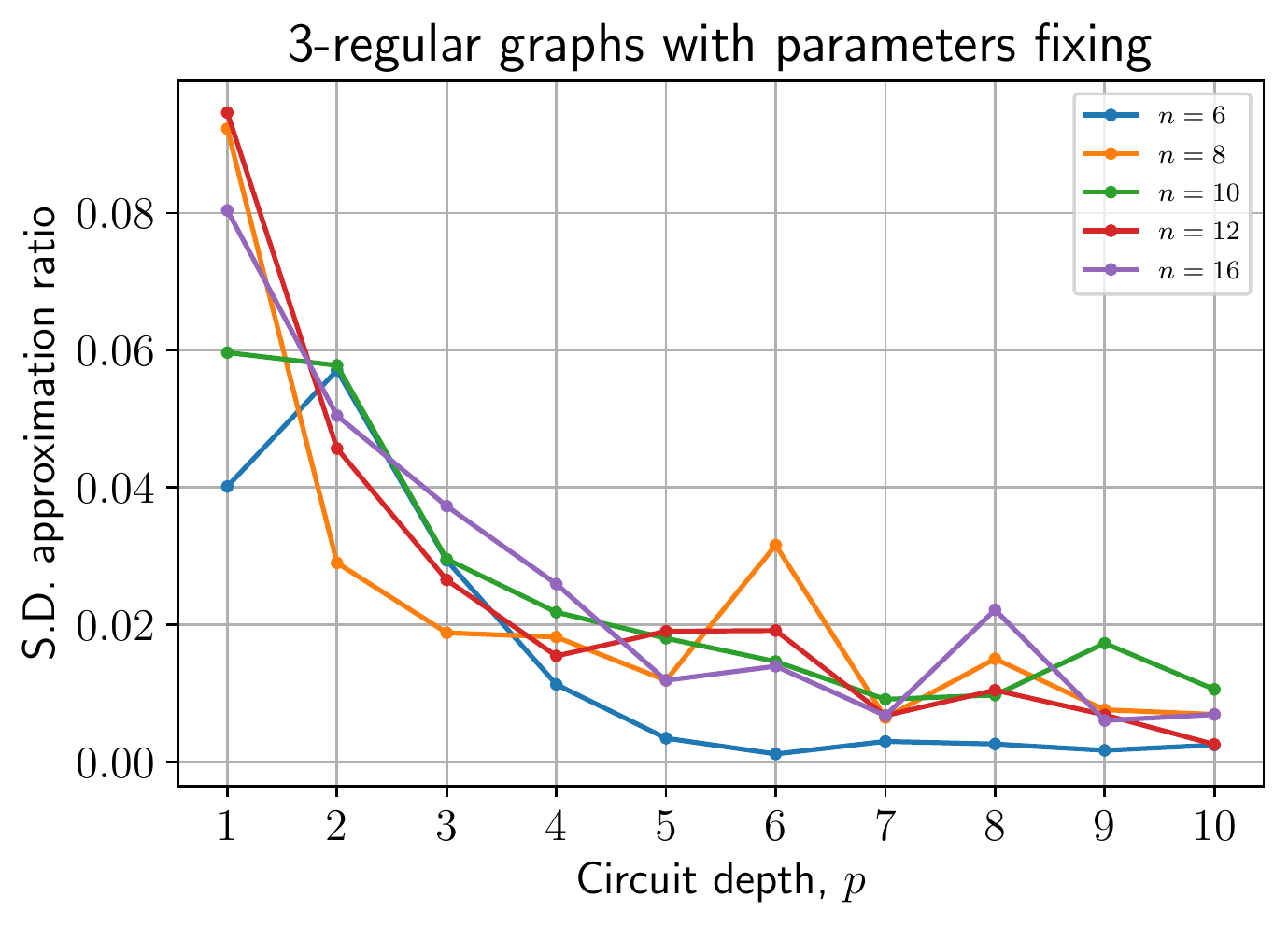}
	\caption{The variation of the standard deviation of the approximation ratio $\alpha$ against the circuit depth $p$, for the 3-regular graphs. This is the same set of approximation ratio as in %
	Fig.~\ref{fig:qaoa-results}(c).}
	\label{fig:reg3-pf-std}
\end{figure}

Fig.~\ref{fig:qaoa-results}(c) and (d) shows the results of the QAOA simulation with the parameters fixing strategy applied. The increasing trend of the mean approximation ratio against the
circuit depth can be observed for both the 3-regular graphs and the \ER graphs. The 3-regular instances are able to achieve mean $\alpha>0.95$ at $p\geq 7$ and the \ER instances achieves mean $\alpha>0.95$
at $p\geq 8$. This is in contrast to the random initialization method, which the mean $\alpha$ does not increase with increasing $p$. Since the mean $\alpha$ given by fixing parameters is relatively high,
this means that the optimizer rarely converges to a false maxima. By starting QAOA with the optimal parameters from shallower depths, the problem of
QAOA converging to a false local minima can thus be subdued, as shown in the results. Hence, we can conclude that this method is effective in obtaining good results in QAOA, at least true for the graph
instances used in our work. Moreover, we also found that for the 3-regular graphs with parameters fixing, the standard deviation of the $\alpha$'s decreases with increasing $p$ as shown in Fig.~\ref{fig:reg3-pf-std}.
This implies that as $p$ increases, the significance of the new random parameters $\Gb{i}$ becomes less, as no matter what the random value is, the value of $\alpha$ does not change much.
The dependence of $\alpha$ is stronger in those optimal parameters from the shallower depths: $(\gamma^*_1,...,\gamma^*_{p-1},\beta^*_1,...,\beta^*_{p-1})$. However, we do not observe this pattern
in the results of the \ER graphs.

\begin{figure}
	\centering
	\includegraphics[width=0.45\textwidth]{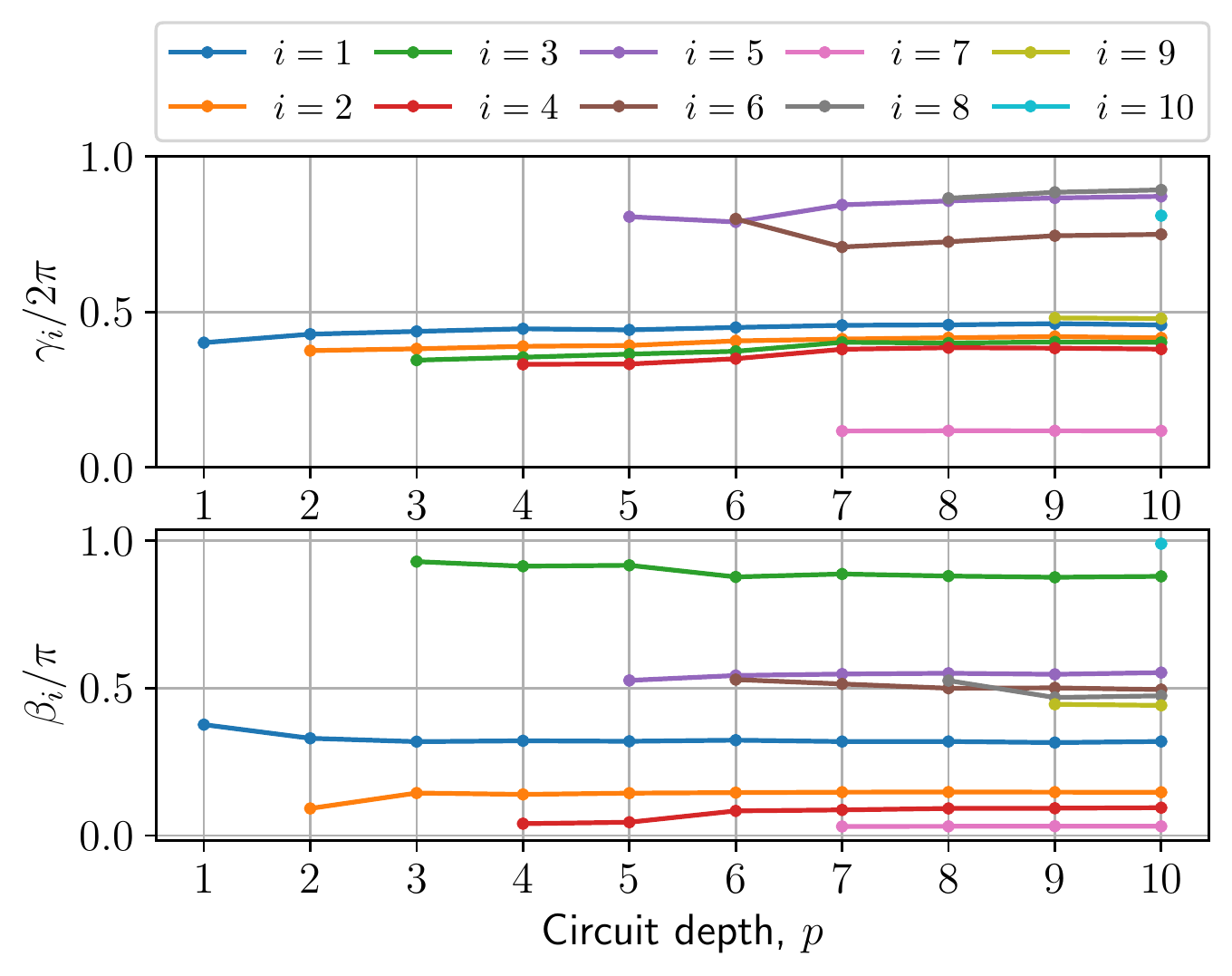}
	\caption{Changes in the $i$-th position optimal parameters as they are passed into the next depth of QAOA as the initial parameters. This is extracted from the simulation data of %
	a 8-node 3-regular graph.}
	\label{fig:var-params}
\end{figure}

Although we call this the parameters fixing strategy, in our actual simulation, we allow the optimal parameters $(\gamma^*_1,...,\gamma^*_{p-1},\beta^*_1,...,\beta^*_{p-1})$ to be further optimized
as they are passed into QAOA of depth $p$ as the initial parameters. Hence, the optimal parameters at $p$ will be $(\gamma^{**}_1,...,\gamma^{**}_{p-1},\gamma^*_p,\beta^{**}_1,...,\beta^{**}_{p-1},\beta^*_p)$,
where the optimal values at $p-1$: $(\gamma^*_i,\beta^*_i)$, is different from the optimal values at $p$: $(\gamma^{**}_i,\beta^{**}_i)$. However, our simulation data in Fig.~\ref{fig:var-params} shows that this
difference is actually subtle. As $p$ increases, the optimal parameter at the $i$-th position stays almost the same relative to other optimal parameters. It is also seen that the parameters tend to
change more when they are first introduced, e.g. $\gamma_6$ at $p=6$, and change less as the depth increases. This shows that the optimal parameters at $p$ stay optimum at $p+1$. If this was not the case,
the values of the optimal parameters at $p$ would have gone through more drastic changes in their values.


\section{Optimization Landscape}
\begin{figure}[t]
	\centering
	\begin{subfigure}[t]{0.45\textwidth}
		\centering
		\sidecaption{subfig:a}
		\raisebox{-\height}{\includegraphics[width=\textwidth]{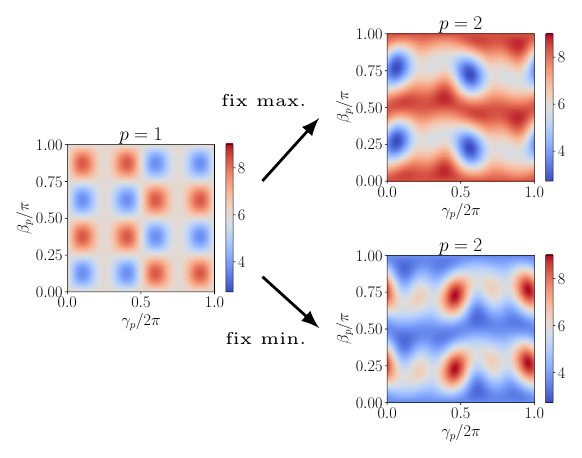}}
	\end{subfigure}
	\begin{subfigure}[t]{0.35\textwidth}
		\centering
		\sidecaption{subfig:b}
		\raisebox{-\height}{\includegraphics[width=\textwidth]{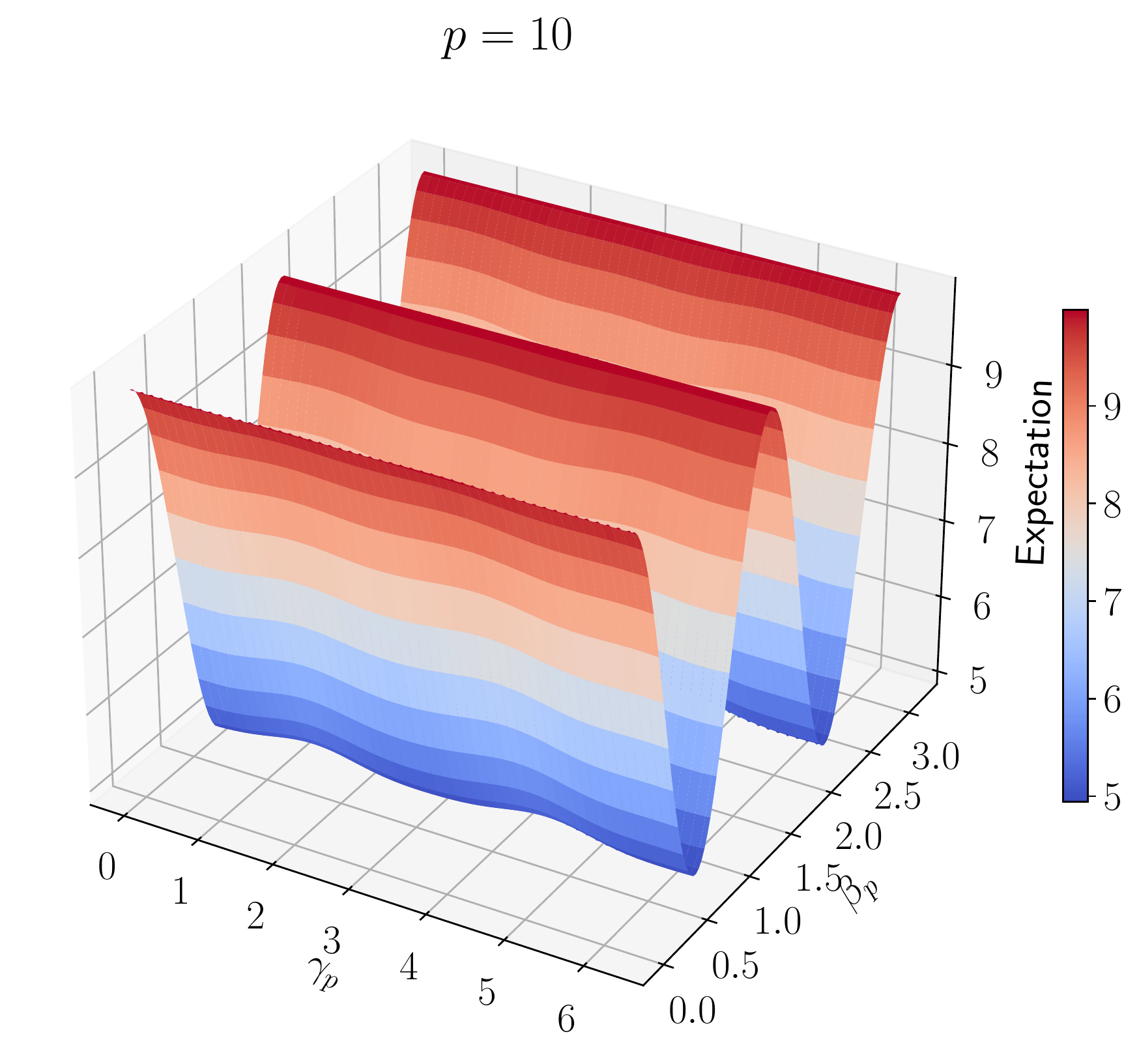}}
	\end{subfigure}
	\caption{(a) Heatmap showing the changes in the optimization landscape when different parameters are fixed while the depth increases. Red regions correspond to high expectation values and blue regions %
		correspond to low expectation values. On the left is the heatmap of the optimization landscape for $p=1$ in the $(\gamma_1,\beta_1)$ space. %
		On the right is the heatmap of $p=2$ in the $(\gamma_2, \beta_2)$ space, while $(\gamma_1,\beta_1)$ is fixed. %
		(b) The landscape of $p=10$ with parameters fixing in the $(\gamma_{10},\beta_{10})$ space. The optimal parameters up until $p=9$ are extracted from real simulation data.
		These are the optimization landscapes of the 8-node 3-regular graph used in our simulation.}
	\label{fig:hm-fix}
\end{figure}

The optimization landscape (the landscape of the expectation function $F_p\Gbvec$) is the landscape traversed by the optimizer. It is a crucial part in QAOA as its properties decide the convergence
of the local optimizers as mentioned in the previous sections. We investigate the optimization landscape to find out the reason behind the effectiveness of the parameters fixing
strategy. We find out that as we fix different points in depth $p$, the optimization landscape of depth $p+1$ is different, depending on which points are fixed. Fig.~\ref{fig:hm-fix}(a) shows
different effects of fixing the parameters of the maximum points (high expectation value) and fixing the parameters of the minimum points (low expectation value) as the depth increases.
When the parameters of the maximum points are fixed, the optimization landscape is covered with more red region, which correspond to high expectation values.
On the other hand, when we fix the parameters of the minimum points, the optimization landscape is covered with more blue region, which correspond to low expectation values. Additionally,
to observe the effect of parameters fixing at large depths, we extract the optimal parameters at $p=10$ from our simulation data and plot the optimization landscape they span. Fig.~\ref{fig:hm-fix}(b)
shows the landscape after parameters fixing is applied continuously until $p=10$. The maximum points in the landscape slowly transforms into maximum lines with increasing depth while fixing the optimal
parameters. We observe similar pattern in the transformation of landscapes in other graph instances as well. We hypothesize that the maximum lines causes the optimizer becomes less likely to be trapped
inside a local maxima. Hence, the approximation ratio is higher on average with the parameters fixing strategy applied.


\section{Discussion and Conclusion}
We have introduced the parameters fixing strategy as an initialization method for QAOA and have shown its effectiveness in practice. The strategy reuses optimal parameters from
the previous depths to obtain high approximation ratio on average at large QAOA circuit depths, which has high practical importance. We show the effectiveness of this strategy using QAOA simulations to solve
the Max-cut problem for the 3-regular graphs and the \ER graphs, and compare the results with those obtained from random initialization. It is shown that with this strategy applied, the average approximation
ratios are higher than those which are initialized randomly. For all the 3-regular instances we considered, the mean approximation ratio $\alpha>0.95$ is achieved when $p\geq 7$. On the other hand, the \ER
instances achieves mean $\alpha>0.95$ when $p\geq 8$. Our results also show that for 3-regular graphs, the significance of new parameters introduced at higher depths becomes less when parameters fixing
is applied. This implies that if the parameters up until depth $p$ are optimal, no matter what new values we choose for $p+1$, the value of $\alpha$ does not change much. Also, we have shown how the optimization
landscape changes as the circuit depth increases during parameters fixing, transforming from maximum points to maximum lines as $p$ increases.

Although the parameters fixing strategy shows good performance in yielding high approximation ratios, it also has apparent drawbacks. As shown in Algorithm~\ref{alg:pf}, this strategy requires
$\mathcal{O}(np)$ time, where $n$ is the number of trials to obtain the maximum expectation, and $p$ is the circuit depth for QAOA to be solved. This strategy forces the user to start solving QAOA
from small depths like $p=1$. However, some has proposed the analytical solution for QAOA at small depths for the 3-regular graphs~\cite{Wurtz_2021,akshay2021parameter}. Therefore, it is possible to obtain
the optimal parameters for small depths analytically. Also, the performance of our strategy is tested only with the NELDER-MEAD optimizer. The performances of the other optimizers are yet to be studied.

It is inevitable to solve QAOA at large circuit depths, as large problem instances usually requires the circuit depth to be increased to obtain a good approximation to the solution. Since local optimizers
perform badly at large circuit depths, a workaround is needed for the optimizers to converge to the desired global maxima.
Although more rigorous proofs are required, the strategy we propose is indeed practically useful, as we have demonstrated its performance using simulation results.
We aim to obtain more rigorous proofs for the parameters fixing strategy, and extend our results to other classes of graphs, or even other problems that can be solved using QAOA.

\bibliographystyle{IEEEtran}
\bibliography{references}

\begin{thebibliography}{10}
\providecommand{\url}[1]{#1}
\csname url@samestyle\endcsname
\providecommand{\newblock}{\relax}
\providecommand{\bibinfo}[2]{#2}
\providecommand{\BIBentrySTDinterwordspacing}{\spaceskip=0pt\relax}
\providecommand{\BIBentryALTinterwordstretchfactor}{4}
\providecommand{\BIBentryALTinterwordspacing}{\spaceskip=\fontdimen2\font plus
\BIBentryALTinterwordstretchfactor\fontdimen3\font minus
  \fontdimen4\font\relax}
\providecommand{\BIBforeignlanguage}[2]{{%
\expandafter\ifx\csname l@#1\endcsname\relax
\typeout{** WARNING: IEEEtran.bst: No hyphenation pattern has been}%
\typeout{** loaded for the language `#1'. Using the pattern for}%
\typeout{** the default language instead.}%
\else
\language=\csname l@#1\endcsname
\fi
#2}}
\providecommand{\BIBdecl}{\relax}
\BIBdecl

\bibitem{farhi2014quantum}
E.~Farhi, J.~Goldstone, and S.~Gutmann, ``A quantum approximate optimization
  algorithm,'' 2014.

\bibitem{lloyd2018quantum}
S.~Lloyd, ``Quantum approximate optimization is computationally universal,''
  2018.

\bibitem{farhi2019quantum}
E.~Farhi and A.~W. Harrow, ``Quantum supremacy through the quantum approximate
  optimization algorithm,'' 2019.

\bibitem{AuriacOptimCuts}
\BIBentryALTinterwordspacing
J.~C.~A. D'Auriac, M.~Preissmann, and A.~Seb\"{o}, ``Optimal cuts in graphs and
  statistical mechanics,'' \emph{Math. Comput. Model.}, vol.~26, no. 8–10, p.
  1–11, Oct. 1997. [Online]. Available:
  \url{https://doi.org/10.1016/S0895-7177(97)00195-7}
\BIBentrySTDinterwordspacing

\bibitem{Kar72}
R.~Karp, ``Reducibility among combinatorial problems,'' in \emph{Complexity of
  Computer Computations}, R.~Miller and J.~Thatcher, Eds.\hskip 1em plus 0.5em
  minus 0.4em\relax Plenum Press, 1972, pp. 85--103.

\bibitem{GW1995}
\BIBentryALTinterwordspacing
M.~X. Goemans and D.~P. Williamson, ``Improved approximation algorithms for
  maximum cut and satisfiability problems using semidefinite programming,''
  \emph{J. ACM}, vol.~42, no.~6, p. 1115–1145, Nov. 1995. [Online].
  Available: \url{https://doi.org/10.1145/227683.227684}
\BIBentrySTDinterwordspacing

\bibitem{Wurtz_2021}
\BIBentryALTinterwordspacing
J.~Wurtz and P.~Love, ``Maxcut quantum approximate optimization algorithm
  performance guarantees for $p>1$,'' \emph{Physical Review A}, vol. 103,
  no.~4, Apr 2021. [Online]. Available:
  \url{http://dx.doi.org/10.1103/PhysRevA.103.042612}
\BIBentrySTDinterwordspacing

\bibitem{Guerreschi_2019}
\BIBentryALTinterwordspacing
G.~G. Guerreschi and A.~Y. Matsuura, ``Qaoa for max-cut requires hundreds of
  qubits for quantum speed-up,'' \emph{Scientific Reports}, vol.~9, no.~1, May
  2019. [Online]. Available: \url{http://dx.doi.org/10.1038/s41598-019-43176-9}
\BIBentrySTDinterwordspacing

\bibitem{niu2019}
M.~Y. Niu, S.~Lu, and I.~L. Chuang, ``Optimizing qaoa: Success probability and
  runtime dependence on circuit depth,'' 2019.

\bibitem{Willsch_2020}
\BIBentryALTinterwordspacing
M.~Willsch, D.~Willsch, F.~Jin, H.~De~Raedt, and K.~Michielsen, ``Benchmarking
  the quantum approximate optimization algorithm,'' \emph{Quantum Information
  Processing}, vol.~19, no.~7, Jun 2020. [Online]. Available:
  \url{http://dx.doi.org/10.1007/s11128-020-02692-8}
\BIBentrySTDinterwordspacing

\bibitem{Moussa_2020}
\BIBentryALTinterwordspacing
C.~Moussa, H.~Calandra, and V.~Dunjko, ``To quantum or not to quantum: towards
  algorithm selection in near-term quantum optimization,'' \emph{Quantum
  Science and Technology}, vol.~5, no.~4, p. 044009, Oct 2020. [Online].
  Available: \url{http://dx.doi.org/10.1088/2058-9565/abb8e5}
\BIBentrySTDinterwordspacing

\bibitem{leo2020}
\BIBentryALTinterwordspacing
L.~Zhou, S.-T. Wang, S.~Choi, H.~Pichler, and M.~D. Lukin, ``Quantum
  approximate optimization algorithm: Performance, mechanism, and
  implementation on near-term devices,'' \emph{Phys. Rev. X}, vol.~10, p.
  021067, Jun 2020. [Online]. Available:
  \url{https://link.aps.org/doi/10.1103/PhysRevX.10.021067}
\BIBentrySTDinterwordspacing

\bibitem{Cook2020TheQA}
J.~Cook, S.~Eidenbenz, and A.~B{\"a}rtschi, ``The quantum alternating operator
  ansatz on maximum k-vertex cover,'' \emph{2020 IEEE International Conference
  on Quantum Computing and Engineering (QCE)}, pp. 83--92, 2020.

\bibitem{Daniel_bangbang}
\BIBentryALTinterwordspacing
D.~Liang, L.~Li, and S.~Leichenauer, ``Investigating quantum approximate
  optimization algorithms under bang-bang protocols,'' \emph{Phys. Rev.
  Research}, vol.~2, p. 033402, Sep 2020. [Online]. Available:
  \url{https://link.aps.org/doi/10.1103/PhysRevResearch.2.033402}
\BIBentrySTDinterwordspacing

\bibitem{sack2021quantum}
S.~H. Sack and M.~Serbyn, ``Quantum annealing initialization of the quantum
  approximate optimization algorithm,'' 2021.

\bibitem{multistart}
\BIBentryALTinterwordspacing
R.~Shaydulin, I.~Safro, and J.~Larson, ``Multistart methods for quantum
  approximate optimization,'' \emph{2019 IEEE High Performance Extreme
  Computing Conference (HPEC)}, Sep 2019. [Online]. Available:
  \url{http://dx.doi.org/10.1109/HPEC.2019.8916288}
\BIBentrySTDinterwordspacing

\bibitem{brandao2018fixed}
F.~G. S.~L. Brandao, M.~Broughton, E.~Farhi, S.~Gutmann, and H.~Neven, ``For
  fixed control parameters the quantum approximate optimization algorithm's
  objective function value concentrates for typical instances,'' 2018.

\bibitem{Alam2020ML}
M.~Alam, A.~Ash-Saki, and S.~Ghosh, ``Accelerating quantum approximate
  optimization algorithm using machine learning,'' in \emph{Proceedings of the
  23rd Conference on Design, Automation and Test in Europe}, ser. DATE
  '20.\hskip 1em plus 0.5em minus 0.4em\relax San Jose, CA, USA: EDA
  Consortium, 2020, p. 686–689.

\bibitem{farhi:qaa}
E.~Farhi, J.~Goldstone, S.~Gutmann, and M.~Sipser, ``Quantum computation by
  adiabatic evolution,'' 2000.

\bibitem{Crooks2018PerformanceOT}
G.~Crooks, ``Performance of the quantum approximate optimization algorithm on
  the maximum cut problem,'' \emph{arXiv: Quantum Physics}, 2018.

\bibitem{aposmm1}
J.~Larson and S.~M. Wild, ``Asynchronously parallel optimization solver for
  finding multiple minima,'' \emph{Mathematical Programming Computation},
  vol.~10, no.~3, pp. 303--332, 2018.

\bibitem{aposmm2}
------, ``A batch, derivative-free algorithm for finding multiple local
  minima,'' \emph{Optimization and Engineering}, vol.~17, pp. 205--228, 2016.

\bibitem{bobyqa}
M.~Powell, ``The bobyqa algorithm for bound constrained optimization without
  derivatives,'' \emph{Technical Report, Department of Applied Mathematics and
  Theoretical Physics}, 01 2009.

\bibitem{cobyla}
\BIBentryALTinterwordspacing
M.~J.~D. Powell, \emph{A Direct Search Optimization Method That Models the
  Objective and Constraint Functions by Linear Interpolation}, 1994, pp.
  51--67. [Online]. Available:
  \url{https://app.dimensions.ai/details/publication/pub.1046127469}
\BIBentrySTDinterwordspacing

\bibitem{neldermead}
\BIBentryALTinterwordspacing
J.~A. Nelder and R.~Mead, ``{A Simplex Method for Function Minimization},''
  \emph{The Computer Journal}, vol.~7, no.~4, pp. 308--313, 01 1965. [Online].
  Available: \url{https://doi.org/10.1093/comjnl/7.4.308}
\BIBentrySTDinterwordspacing

\bibitem{akshay2021parameter}
V.~Akshay, D.~Rabinovich, E.~Campos, and J.~Biamonte, ``Parameter concentration
  in quantum approximate optimization,'' 2021.

\end{thebibliography}

\end{document}